\documentclass[12pt]{iopart}

\usepackage{graphicx}

\usepackage{iopams}
\usepackage{amsfonts}

\begin{document}

\title[Dirac particle radiation]
{Classical Dirac particle: Mass and Spin invariance and radiation reaction}

\author{Mart\'{\i}n Rivas}
\address{Theoretical Physics Department, The University of the Basque Country,\\ 
Bilbao, Spain}
\ead{martin.rivas@ehu.eus}

\begin{abstract}
According to the atomic principle \cite{atomic} an elementary particle has no excited states and under any interaction, if it is not annihilated, its internal structure cannot be modified. The intrinsic properties are the mass $m$ and the absolute value of the spin in the center of mass frame $S=\hbar/2$. We analyze the closed system made of a single Dirac particle and an external electromagnetic field. The Poincar\'e invariance of the dynamics implies that the energy, linear momentum and angular momentum of the whole system must be conserved. The Dirac particle has two distinguished points, the center of charge ${\bi r}$ and the center of mass ${\bi q}$. When interacting, the energy expended by the field is the work done by the external Lorentz force along the center of charge trajectory. The variation of the mechanical energy of the particle is the work done by the external Lorentz force along the center of mass trajectory. If these two works are different, the excess of energy must be transformed into radiation, returning that energy to the field. The accelerated Dirac particle radiates. Accelerated spinless particles do not radiate. We analyze the spin dynamics of the Dirac particle under an external electromagnetic field.  The requirement that the absolute value of the spin for the center of mass observer cannot be modified by the interaction, give rise to a twofold feature: a modification of the dynamical equation of the acceleration of the CM, which includes a new reaction force along the linear momentum that suggests a linear momentum emission in the opposite direction, and also an external reaction torque, that can be interpreted as the radiation of angular momentum in the orthogonal direction to the zitter plane.
\end{abstract}

%%%%%%%%%%%%%%%%%
\section{Introduction}
\label{Introduction}
%%%%%%%%%%%%%%%%%
The classical description of a Dirac particle is obtained from a general formalism for describing elementary
spinning particles \cite{Rivasbook}. In the recent publication \cite{Beck} Beck shows the equivalence and similitudes of  formulations by Barut and Zanghi (1984) \cite{Barut}, Salesi (2002) \cite{Salesi}, 
Hestenes (2010) \cite{Hestenes}, Beck (2023) \cite{Beck2} and our formulation \cite{Rivasbook} for describing the electron.
In the article \cite{ClassicalDiracI} we offer a simplified and detailed version of the formalism where the reader can find the classical description of the Dirac particle, the dynamical equations in the presence of an external electromagnetic field, and the definition of the fundamental observables, energy, linear momentum and spins, as well as links to Wolfram's Mathematica Notebooks for solving several examples of the interaction of the Dirac particle with external electric and magnetic fields. The expressions of the different observables are obtained by means Noether's theorem. 

The difference with Beck's formalism \cite{Beck}, is that these two points which he calls ``center of inertia" and ``center of spin", respectively, are considered with simultaneity only in the proper frame where the center of inertia is at rest, while their Poincar\'e transformed positions to another inertial reference frame are considered at different times, because synchronous events in one frame are not synchronous in another. In our formalism the CM is defined at every inertial reference frame at the same time than the CC, it is expressed in terms of the CC and their time derivatives and therefore their trajectories are computed and depicted simultaneously.

The main feature is that the classical Dirac particle is described by a single point ${\bi r}$, interpreted as the center of charge (CC) of the particle, moving at the speed of light, and that satisfies a system of ordinary differential equations of fourth order. These equations are obtained as the Euler-Lagrange equations of the mechanical system. All observables can thus be expressed in terms of the point ${\bi r}$ and the different time derivatives. One consequence of the Noether analysis is that a point ${\bi q}$, different than ${\bi r}$, can be defined at the same time $t$ and interpreted as the center of mass of the particle (CM). With this definition, the fourth order system of differential equations can be transformed into a coupled system of second order differential equations for the variables ${\bi r}$ and ${\bi q}$. The evolution parameter is the time coordinate $t$ in the corresponding inertial reference frame, and the two positions ${\bi r}$ and ${\bi q}$ are considered simultaneously as functions of $t$.  This allows us to obtain two different angular momentum observables: the spin with respect to the CC, ${\bi S}_{CC}$, and  the spin with respect to the CM, ${\bi S}_{CM}$
that can be expressed in terms of both points and their corresponding velocities. Both spins satisfy different dynamical equations and the CC spin ${\bi S}_{CC}$, satisfies the same dynamical equation as Dirac's spin operator in the quantum case.

In the article \cite{atomic} we considered the Atomic Principle as a fundamental principle which states that an elementary particle does not have excited states and, if it is not annhilated, its internal structure cannot be modified by any interaction. The analysis is done in a natural system of units, where $\hbar=c=1$, and the mass and charge of the electron $m=e=1$, the unit of length is $2R_0=\hbar/mc=1$, and the unit of time is $\tau=2R_0/c$, where $R_0$ is the separation between the CC and the CM for the center of mass observer.

We start in section {\bf\ref{spindynamiccs}} the analysis of the dynamical equations of both spins, in order to obtain the condition that the absolute value of the spin for the center of mass observer cannot be modified by any interaction.
This leads in Section {\bf\ref{InvSpin}} to the modification of the Euler-Lagrange equations where a reaction force is defined. By assuming that the mass of the particle remains invariant, suggests that not all the energy expended by the field is transformed into mechanical energy of the particle. In Sections {\bf\ref{difpradiado}} and {\bf\ref{difHradiado}} we analyze the radiation of linear momentum and energy, respectively. We finally analyze in Sec. {\bf\ref{difspin}} the radiation of angular momentum.

%%%%%%%%%%%%%%%%%
\section{Spin dynamics}
\label{spindynamiccs}
%%%%%%%%%%%%%%%%%
The dynamical equations of the center of mass (CM) ${\bi q}$ and the center of charge (CC) ${\bi r}$ of the Dirac particle under any electromagnetic field obtained from the general Lagrangian $L=L_0+L_{em}$, where $L_0({\bi u},{\bi a},\bomega)$ is the free Lagrangian and $L_{em}=-eA_0(t,{\bi r})+e{\bi u}\cdot{\bi A}(t,{\bi r})$, where ${\bi u}=d{\bi r}/dt$, and in terms of scalar and vector potentials $A_0$ and ${\bi A}$ defined at time $t$ at the center of charge position ${\bi r}$, respectively, are, in the natural system of units \cite{ClassicalDiracI}:
\begin{eqnarray}
\frac{d{\bi q}}{dt}&=&{\bi v},\quad\frac{d{\bi v}}{dt}=\frac{1}{\gamma(v)}\left[{\bi F}-{\bi v}\left({\bi F}
\cdot{\bi v}\right)\right],\quad {\bi F}={\bi E}(t,{\bi r})+{\bi u}\times{\bi B}(t,{\bi r}),\label{eq:d2qdt2}\\
\frac{d{\bi r}}{dt}&=&{\bi u},\quad \frac{d{\bi u}}{dt}=\frac{1-{\bi v}\cdot{\bi u}}{({\bi q}-{\bi r})^2}({\bi q}-{\bi r}),\qquad \gamma(v)=(1-v^2)^{-1/2},\label{eq:d2rdt2}
 \end{eqnarray}
with the constraints $u=1$, $v<1$. In the derivation of these dynamical equations we have made no use of a possible radiation reaction force.

Equation (\ref{eq:d2qdt2}) is $d{\bi p}_m/dt={\bi F}$, where ${\bi p}_m=\gamma(v)m{\bi v}$ is the mechanical linear momentum, and equation (\ref{eq:d2rdt2}) comes from the definition of the CM position
\[
{\bi q}={\bi r}+\frac{1-{\bi v}\cdot{\bi u}}{a^2}{\bi a},\quad {\bi a}=\frac{d{\bi u}}{dt}.
\]

Equation (\ref{eq:d2qdt2}) for an electron in an external field, if the electric field is expressed in the International System of units in V/m, and the magnetic field in Teslas, is written as:
\begin{equation}
\frac{d{\bi v}}{dt}=\frac{K_E}{\gamma(v)}\left[{\bi E}-{\bi v}\left({\bi E}
\cdot{\bi v}\right)\right]+\frac{K_B}{\gamma(v)}\left[{\bi u}\times{\bi B}-{\bi v}\left(({\bi u}\times{\bi B})\cdot{\bi v}\right)\right],
\label{natdifeq}
\end{equation}
with the remaing variables in dimensionless natural units and the constants
\[
K_E=\frac{e\hbar}{m^2c^3}=7.55676\cdot10^{-19}\;{\rm m/V},\quad K_B=\frac{e\hbar}{m^2c^2}=2.26546\cdot10^{-10}\;{\rm T}^{-1},
\]
are the conversion factors from  IS to the natural system of units, and are the inverses of the so-called Schwinger limits for the electric and magnetic field, respectively. 

The general expressions of both spins for the Dirac particle with respect to ${\bi r}$ and ${\bi q}$, respectively, in natural units, are
\begin{equation}
{\bi S}_{CC}=-\gamma(v)({\bi r}-{\bi q})\times{\bi u},
\label{sCC}
\end{equation}
\begin{equation}
{\bi S}_{CM}={\bi S}_{CC}+({\bi r}-{\bi q})\times{\bi p}_m=-{\gamma(v)}({\bi r}-{\bi q})\times({\bi u}-{\bi v}),
\label{sCM}
\end{equation}
where the variables ${\bi r}$, ${\bi q}$, ${\bi u}$ and ${\bi v}$ and the mechanical linear momentum ${\bi p}_m=\gamma(v){\bi v}$, and mechanical energy $H_m=\gamma(v)$, are expressed in natural units. In general, the absolute value of the CM spin will be a function of the center of mass velocity ${\bi v}$, $S_{CM}(v)$,
with the value at rest  $S_{CM}(0)=1/2$. The dynamics modifies the variables they depend and their actual value will be determined during the dynamical process. Because the dynamical equation (\ref{eq:d2rdt2}) the acceleration of the CC has the direction of the relative vector $({\bi r}-{\bi q})$, so that we see from (\ref{sCC}) that the CC spin ${\bi S}_{CC}$ is always orthogonal to the plane subtended by the vectors ${\bi u}$ and $({\bi r}-{\bi q})$. We call this plane the zitterbewegung plane of the particle.

Taking the time derivatives of (\ref{sCC}) and (\ref{sCM}) we get in the free case, where ${\bi v}$ is constant:
\[
 \frac{d{\bi S}_{CC}}{dt}={\bi p}_m\times{\bi u},\quad \frac{d{\bi S}_{CM}}{dt}=0,
\]
respectively, 
and the CM spin ${\bi S}_{CM}$ is conserved but the spin ${\bi S}_{CC}$ satisfies the same dynamical equation than Dirac's spin operator in the quantum case. It is the classical spin equivalent of Dirac's spin operator.

The total angular momentum of the particle ${\bi J}$, with respect to the origin of the observer's frame can be written in two different ways:
\[
{\bi J}={\bi r}\times{\bi p}_m+{\bi S}_{CC},\quad {\bi J}={\bi q}\times{\bi p}_m+{\bi S}_{CM}.
\]
If ${\bi F}_T$ is the total external force defined at the CC position ${\bi r}$, the time variation of ${\bi J}$ will be the torque of this force with respect to the origin:
\[
\frac{d{\bi J}}{dt}={\bi r}\times{\bi F}_T.
\]
From the above two expressions of ${\bi J}$ we arrive, under any interaction to
\begin{equation}
 \frac{d{\bi S}_{CC}}{dt}={\bi p}_m\times{\bi u},\quad \frac{d{\bi S}_{CM}}{dt}=({\bi r}-{\bi q})\times{\bi F}_T.
\label{torques}
\end{equation}
The CC spin ${\bi S}_{CC}$ satisfies the same spin dynamical equation than Dirac's spin operator even under interaction, while the CM spin dynamical equation is the torque of the external force defined at CC point ${\bi r}$, with respect to the CM point ${\bi q}$.

%%%%%%%%%%%%%%%%%
\section{The atomic principle}
\label{atomic}
\begin{figure}[!hbtp]\centering%
 \includegraphics[width=6.5cm]{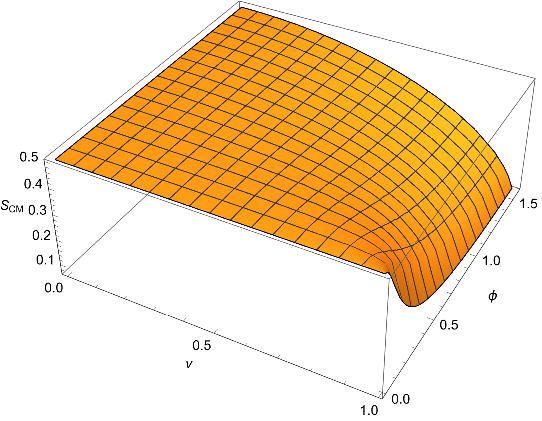}\hspace{1cm} \includegraphics[width=4cm]{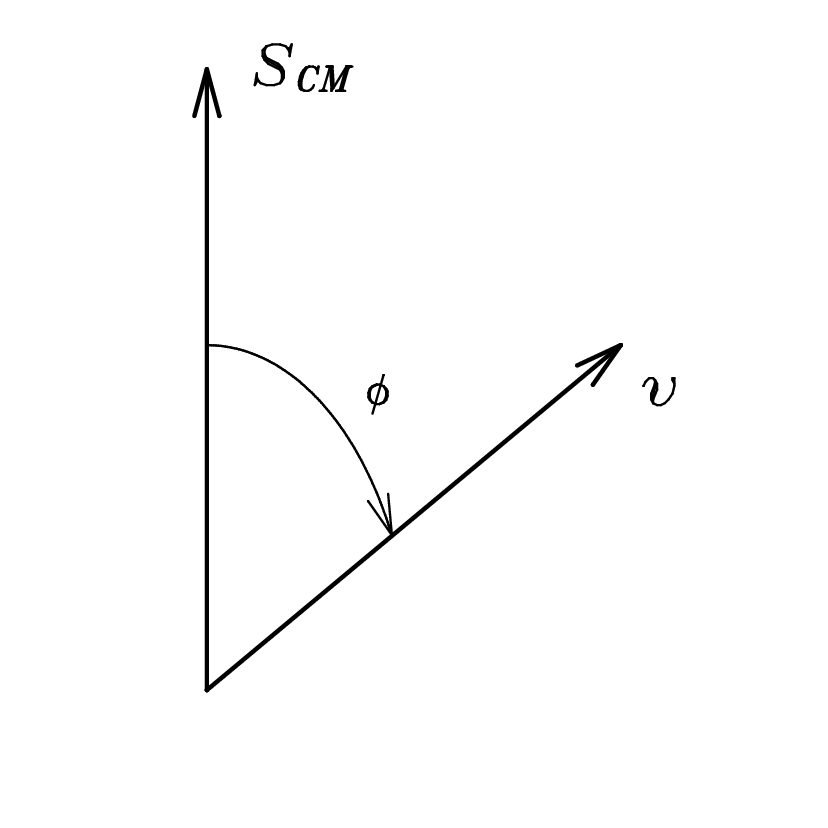}
\caption{Absolute value of the CM spin ${\bi S}_{CM}({\bi v})$, $S_{CM}(v,\phi)$, given in (\ref{ModSpinCM}) in any inertial reference frame, as a function of the CM velocity $v\in[0,1)$ and the angle $\phi\in[0,\pi/2]$ between the CM velocity ${\bi v}$ and the CM spin ${\bi S}_{CM}({\bi v})$.}  
\label{variation}
\end{figure}

%%%%%%%%%%%%%%%%%
The assumption of the Atomic Principle \cite{atomic} implies that the internal structure of an elementary particle cannot be modified by any interaction. The intrinsic properties of an elementary particle have to remain the same. The intrinsic properties of the Dirac particle are related to the two invariants or Casimir operators of the Poincar\'e group. They are the absolute value of the two four-vectors, the energy-momentum $p^{\mu}$ and the Pauli-Lubanski $w^\mu=(1/2)\epsilon^{\mu\nu\sigma\rho}p_\nu J_{\sigma\rho}$ four-vectors, orthogonal to each other $p^\mu w_\mu=0$. They are expressed in terms of the mechanical energy and mechanical linear momentum $H_m$ and ${\bi p}_m$, respectively,  and the CM spin ${\bi S}_{CM}$ as \cite{ClassicalDiracI}:
\[
p^\mu\equiv(H_m/c,{\bi p}_m)=(\gamma(v)mc,\gamma(v)m{\bi v}),\quad w^\mu\equiv({\bi p}_m\cdot{\bi S}_{CM},H_m{\bi S}_{CM}/c),
\]
\[
p^\mu\equiv\gamma(v)(mc,m{\bi v}),\quad w^\mu\equiv\gamma(v)\left(m{\bi v}\cdot{\bi S}_{CM},mc{\bi S}_{CM}\right).
\]
and the invariant properties in natural units,  are:
\[
p^\mu p_\mu=m^2c^2=1,\quad w^\mu w_\mu=-m^2c^2 S_{CM}(0)^2=-1/4,\quad p^\mu w_\mu=0,
\]
where $m=1$ is the rest mass and $S_{CM}(0)=1/2$, the absolute value of the spin in the center of mass frame, in natural units. From the invariant and constant of the motion $w^\mu({\bi v}) w_\mu({\bi v})=w^\mu(0) w_\mu(0)=-1/4$, in natural units, we get
\[
\gamma^2v^2S_{CM}({\bi v})^2\cos^2\phi-\gamma^2v^2S_{CM}({\bi v})^2=-\frac{1}{4},
\]
and the absolute value of the CM spin for an arbitrary inertial observer is a function of the value of the CM velocity $v$ and the angle $\phi$ between ${\bi S}_{CM}$ and ${\bi v}$, given by:
\begin{equation}
S_{CM}(v,\phi)=\frac{1}{2}\sqrt{\frac{1-v^2}{1-v^2\cos^2\phi}}\;\le \frac{1}{2},
\label{ModSpinCM}
\end{equation}
expression which represented in the figure {\bf\ref{variation}}.

We distinguish between the total energy of a particle $H=H_m+H_I$ where the mechanical energy of the particle is $H_m=\gamma(v)mc^2$, and $H_I$ is the interacting energy. Similarly, the total linear momentum of a particle is
${\bi p}={\bi p}_m+{\bi p}_I$, where the mechanical linear momentum ${\bi p}_m=\gamma(v)m{\bi v}$ and ${\bi p}_I$ the interacting linear momentum. With the minimal coupling Lagrangian $H_I=eA_0$, ${\bi p}_I=e{\bi A}$, in terms of the scalar and vector potentials, respectively. These observables always satisfy the invariant equation:
\[
H_m^2-{\bi p}_m^2=(H-eA_0)^2-({\bi p}-e{\bi A})^2=m^2c^2.
\]

%%%%%%%%%%%%%%%%%
\subsection{Invariance of spin}
\label{InvSpin}
%%%%%%%%%%%%%%%%%
The dynamical equation (\ref{eq:d2qdt2}) is $d{\bi p}_m/dt={\bi F}={\bi E}+{\bi u}\times{\bi B}$ and comes from the minimal coupling interaction electromagnetic Lagrangian, so that the external force that produces that acceleration of the CM is the external Lorentz force. No other condition is required in its deduction. Nevertheless the Atomic Principle requires that the absolute value of the spin for the center of mass observer and the mass of the particle are invariant properties of value $1/2$ and $1$, respectively, in natural units. This fact is not fulfilled in that equation. Although the spin is of constant absolute value for the center of mass observer, the time derivative of ${\bi S}_{CM}(0)$ will be, in general, different from zero. But if the absolute value must remain constant this implies that this time derivative must be orthogonal to the spin vector in this frame. The external force has to produce a torque with respect to the CM, such that the orientation of the spin can change but not the size. The external force is completely arbitrary so that the spin motion must meet this requirement and the end point of the spin vector must move in the orthogonal direction irrespective of the external force. 

The center of mass spin has the form (\ref{sCM}) and its time derivative is given in (\ref{torques}). Let us describe the magnitudes for the center of mass observer with an asterisk. That time derivative in the CM frame looks
\begin{equation}
\frac{d{\bi S}^*_{CM}}{dt^*}={\bi r}^*\times{\bi F}^*,
\label{eq:espinCM}
\end{equation}
since ${\bi q}^*=0$ in this frame. The external force at the CM frame is
\begin{equation}
{\bi F}^*={\bi E}^*+{\bi u}^*\times{\bi B}^*,
\label{dyneqCM}
\end{equation}
in natural units and with the fields evaluated in this frame. If ${\bi u}$ and ${\bi E}$ and ${\bi B}$ are the CC velocity and the electric and magnetic field in the laboratory frame, respectively, these magnitudes for the center of mass observer are:
\begin{equation}
{\bi u}^*=\frac{{\bi u}-\gamma(v){\bi v}+\frac{\gamma^2}{1+\gamma}({\bi v}\cdot{\bi u}){\bi v}}{\gamma(v)(1-{\bi v}\cdot{\bi u})},
\label{transfu}
\end{equation}
\begin{equation}
{\bi E}^*=\gamma(v){\bi E}-\frac{\gamma^2}{1+\gamma}({\bi v}\cdot{\bi E}){\bi v}+\gamma({\bi v}\times{\bi B}).
\label{transfE}
\end{equation}
\begin{equation}
{\bi B}^*=\gamma(v){\bi B}-\frac{\gamma^2}{1+\gamma}({\bi v}\cdot{\bi B}){\bi v}-\gamma({\bi v}\times{\bi E}),
\label{transfB}
\end{equation}
where ${\bi v}$ is the velocity of the center of mass of the Dirac particle in the laboratory frame.
The condition that the absolute value of the spin must be constant in the CM frame is
\[
{\bi S}^*_{CM}\cdot\frac{d{\bi S}^*_{CM}}{dt^*}=0,\quad {\rm where }\quad {\bi S}^*_{CM}=-{\bi r}^*\times{\bi u}^*.
\]
Taking into account (\ref{eq:espinCM}), this scalar product leads to
\[
({\bi r}^*\times{\bi u}^*)\cdot\left({\bi r}^*\times{\bi F}^*\right)={r^*}^2\left({\bi u}^*\cdot{\bi F}^*\right)-({\bi r}^*\cdot{\bi u}^*)\left({\bi r}^*\cdot{\bi F}^*\right),
\] 
so that the external Lorentz force in the center of mass frame (\ref{dyneqCM}) must fulfill
\begin{equation}
{\bi u}^*\cdot{\bi F}^*=0,
\label{radreact}
\end{equation}
since $({\bi r}^*\cdot{\bi u}^*)=0$, because $u=1$. In the center of mass frame the CC velocity must be oriented in the direction orthogonal to the external force. If the external force ${\bi F}^*$ has a component along the CC velocity ${\bi u}$, the torque of this force with respect to the CM will produce a modification of the length of the CM spin ${\bi S}^*_{CM}$.

Making the scalar product of (\ref{transfu}) and (\ref{dyneqCM}) we arrive to:
\begin{equation}
{\bi v}\cdot{\bi E}={\bi u}\cdot{\bi E}+{\bi u}\cdot({\bi v}\times{\bi B}),
\label{radreacCondi1}
\end{equation}
so that by replacing this into the equation (\ref{eq:d2qdt2}) it is transformed into
\begin{equation}
\frac{d{\bi v}}{dt}\Big|_R=\frac{1}{\gamma(v)}\left[{\bi E}+{\bi u}\times{\bi B}-{\bi v}\left({\bi u}\cdot{\bi E}\right)
\right]=\frac{1}{\gamma(v)}\left[{\bi F}-{\bi v}\left({\bi u}\cdot{\bi F}\right)\right].
\label{radreacCondi2}
\end{equation}
The subindex $R$ is for {\it Reaction} or {\it Radiation}.

In the dynamical equation (\ref{eq:d2qdt2}) the CM acceleration is produced by the external Lorentz force ${\bi F}$. It contains a first term which is the Lorentz force modified by a breaking term in the direction to the CM velocity ${\bi v}$. This factor represents the work per unit time of the Lorentz force along the CM displacement $-{\bi v}({\bi v}\cdot{\bi F})$. This term, opposite to the velocity of the CM, prevents the increasing of the CM velocity with no limit, together with the $\gamma(v)$ factor at the denominator. In the high velocity limit the acceleration in both equations (\ref{eq:d2qdt2}) and (\ref{radreacCondi2}) goes to zero. With the consideration of constant absolute value of the spin this braking term has been substituted by a term in (\ref{radreacCondi2}), opposite to the CM velocity as before, but now the intensity is the work of the Lorentz force along the CC displacement $-{\bi v}({\bi u}\cdot{\bi F})$. If the expression of the CM acceleration has changed, this means that the external force has to be different than the Lorentz force that gives rise to the CM acceleration described in equation (\ref{eq:d2qdt2}).

From the electromagnetic point of view, the energy expended by the field when acting on a point charge is the work of the external force along the path of this point charge. From the mechanical point of view, the energy acquired by a massive elementary particle is the work of the external force along the trajectory of the center of mass of the particle, as we discuss in the next subsections.

If we consider the complete system of the external field and the Dirac particle, the total energy, linear momentum and angular momentum must be conserved. If the above two works are different, and since there is no massive particle creation, the difference between those energies represents a modification of the energy of the field. We can interpret this difference as the emission of the extra energy by the Dirac particle in the form of radiation, in the form of energy of the field. From the particle's point of view, the Dirac particle radiates.

%%%%%%%%%%%%%%%%%%%%%%%%%%%%%
\subsection{Radiation of linear momentum}
\label{difpradiado}
%%%%%%%%%%%%%%%%%%%%%%%%%%%%%
The variation of the  mechanical linear momentum without conservation of the spin is the total 
external force, i.e., the Lorentz force, obtained by the usual Lagrangian formalism:
\begin{equation}
\frac{d{\bi p}_m}{dt}={\bi F}.
\label{varpsin}
\end{equation}
When there is conservation of the spin at the CM frame we have to compute de variation of ${\bi p}_{mR}=\gamma(v_R){\bi v}_R$, by using the modified equation of the CM acceleration (\ref{radreacCondi2})
\[
\frac{d{\bi p}_{mR}}{dt}=\frac{d}{dt}(\gamma(v){\bi v})\Big|_R=\gamma(v)^3\left({\bi v}\cdot\frac{d{\bi v}}{dt}\Big|_R\right){\bi v}+\gamma\frac{d{\bi v}}{dt}\Big|_R=
\]
\[
=\gamma^2\left[({\bi v}\cdot{\bi F})-v^2({\bi u}\cdot{\bi F})\right]{\bi v}+{\bi F}-{\bi v}({\bi u}\cdot{\bi F}),
\]
\begin{equation}
\frac{d{\bi p}_{mR}}{dt}={\bi F}-\gamma\left[({\bi u}-{\bi v})\cdot{\bi F}\right]{\bi p}_m,\quad
d{\bi p}_{mR}={\bi F}dt-\gamma[(d{\bi r}-d{\bi q})\cdot{\bi F}]{\bi p}_m,
\label{varpcon}
\end{equation}
since $\gamma^2v^2=\gamma^2-1$, and $d{\bi p}_{mR}=({\bi F}+{\bi F}_R)dt$, where
\begin{equation}
{\bi F}_R=-\gamma[({\bi u}-{\bi v})\cdot{\bi F}]{\bi p}_m,
\label{RadRecF}
\end{equation}

The total external force now, is the original Lorentz force ${\bi F}$, plus a reaction force (\ref{RadRecF}) along the linear momentum ${\bi p}_m$, of magnitude this linear momentum ${\bi p}_m$ multiplied by the difference between the work per unit time of the Lorentz force along the CC trajectory minus the work along the CM trajectory, times the factor $\gamma(v)$. 
The sign of this term depends on the sign of the term $(d{\bi r}-d{\bi q})\cdot{\bi F}$. If this term is positive, the minus sign implies that this force has the direction opposite to ${\bi p}_m$, and thus is a breaking force along the CM trajectory, while if the term $(d{\bi r}-d{\bi q})\cdot{\bi F}$ is negative, this reaction force is along the linear momentum ${\bi p}_m$.
This term (\ref{RadRecF}) is the {\bf reaction force} ${\bi F}_R$, as a consequence of the conservation of the spin for the center of mass observer.

We can interpret the origin of this force in the fact that the particle is radiating linear momentum to the field in the opposite direction. The radiated linear momentum is the opposite of that expression
\begin{equation}
d{\bi p}_R=-{\bi F}_R dt=\gamma[(d{\bi r}-d{\bi q})\cdot{\bi F}]{\bi p}_m.
\label{RadP}
\end{equation}

%%%%%%%%%%%%%%%%%%%%%%%%%%%%%
\subsection{Invariance of the mass. Radiation of energy}
\label{difHradiado}
%%%%%%%%%%%%%%%%%%%%%%%%%%%%%
The invariant $H_m^2-{\bi p}_m^2=1$, in natural units, is such that its time derivative
\[
H_m\frac{dH_m}{dt}-{\bi p}_m\cdot\frac{d{\bi p}_m}{dt}=0, 
\]
and using the expressions of $H_m=\gamma(v)$ and ${\bi p}_m=\gamma(v){\bi v}$ we arrive to
\[
\frac{dH_m}{dt}={\bi v}\cdot\frac{d{\bi p}_m}{dt},
\]
and the variation of the mechanical energy, because of the mass conservation, is
\begin{equation}
{dH_m}=d{\bi q}\cdot{\bi F},
\label{varHsin}
\end{equation}
is the work of the total external force along the CM trajectory, that in this case is the Lorentz force. This is usually known as the {\it center of mass theorem}.

However, the acceleration of the CM when the spin is conserved is (\ref{radreacCondi2}) and the variation of the mechanical energy under the requirement of spin conservation $H_{mR}$, is
\begin{equation}
\frac{dH_{mR}}{dt}=\frac{d\gamma(v)}{dt}\Big|_R=\gamma(v)^3\;{\bi v}\cdot\frac{d{\bi v}}{dt}\Big|_R=\gamma^2({\bi v}\cdot{\bi F})-(\gamma^2-1)({\bi u}\cdot{\bi F}),
\label{vargamma}
\end{equation}
or
\begin{equation}
\frac{dH_{mR}}{dt}={\bi v}\cdot({\bi F}+{\bi F}_R),\quad {dH_{mR}}=d{\bi q}\cdot({\bi F}+{\bi F}_R),
\label{varHcon}
\end{equation}
as the work of the total external force along the CM trajectory, which now is the external Lorentz ${\bi F}$ force plus the reaction force ${\bi F}_R$.

The energy expended by the external field is $dH_f=d{\bi r}\cdot{\bi F}$, i.e., the work of the external Lorentz force along the CC trajectory where the point charge is located. Not all this energy have been used to suply mechanical energy to the particle, and therefore the difference
\begin{equation}
dH_R=dH_f-dH_{mR}=\gamma[(d{\bi r}-d{\bi q})\cdot{\bi F}]H_m,
\label{RadH}
\end{equation}
represents the energy that has returned to the field in the form of radiated energy by the particle. It is proportional to the energy of the particle $H_m$, multiplied by the same radiation factor than the radiated linear momentum (\ref{RadP}).

If the external force tries to modify the absolute value of the spin, the particle reacts by avoiding this modification and emitting or returning electromagnetic energy, linear momentum, as well as different observables, to the external field, and thus producing the above reaction force (\ref{RadRecF}).

Because the trajectories of ${\bi r}$ and ${\bi q}$ are different the conservation of energy, linear and angular momentum imply a modification of the field densities of the above magnitudes in the surrounding of the particle, suggesting that the modification of these densities of the electromagnetic field corresponds to the contribution of the retarded fields of the particle. These retarded fields, when added to the external field, produce a modification of said densities, in order to the conservation laws of the  particle$+$field closed system. This process can be interpreted as a continuous transmission of radiation between the Dirac particle and the field.

%%%%%%%%%%%%%%%%%%%%%%%%%
\section{Radiation of angular momentum}
\label{difspin}
%%%%%%%%%%%%%%%%%%%%%%%%%

We have seen that the dynamical equations for the acceleration of the CM of the Dirac particle (\ref{eq:d2qdt2}), obtained from a minimal coupling interaction Lagrangian with an external electromagnetic field, have been modified by the requirement that the absolute value of the spin for the center of mass observer must remain invariant, and are transformed into (\ref{radreacCondi2}). This implies that the total external force is now the external Lorentz force ${\bi F}$, plus the reaction force ${\bi F}_R$. The work of this reaction force has also contributed to the variation of the mechanical energy of the particle. What we expect is that the torque of this reaction force (\ref{RadRecF}), will also contribute to the dynamics of the CM spin of the particle.
\begin{equation}
\frac{d{\bi S}_{CM}}{dt}=({\bi r}-{\bi q})\times({\bi F}+{\bi F}_R),
\label{torqFFR}
\end{equation}
as the torque of the total external force defined at point ${\bi r}$, with respect to the CM ${\bi q}$.

To compute the variation of the CM spin we have to take the time derivative of the expression (\ref{sCM}), taking into account that for the CM acceleration we have to use the modified dynamical equation (\ref{radreacCondi2}) where the external force that produces this new acceleration is ${\bi F}+{\bi F}_R$. If we take the time derivative of ${\bi S}_{CM}$ we arrive to:
\begin{equation}
\frac{d{\bi S}_{CM}}{dt}=({\bi r}-{\bi q})\times({\bi F}+{\bi F}_R+{\bi F}_S),
\label{torqFFRS}
\end{equation}
to the expected torque equation (\ref{torqFFR}) plus another torque produced by an external force
\[
{\bi F}_S=\gamma(v)^2[({\bi u}-{\bi v})\cdot{\bi F}]{\bi u}, 
\]
so that this new external torque is
\begin{equation}
{\bi M}_S=({\bi r}-{\bi q})\times{\bi F}_S=-\gamma(v)[({\bi u}-{\bi v})\cdot{\bi F}]{\bi S}_{CC}.
\label{torqFS}
\end{equation}
We call this torque ${\bi M}_S$, the {\bf spin reaction torque}. This modification corresponds to a zero external force but a non-zero torque ${\bi M}_S$, and thus no modification of the radiated energy and linear momentum, but the origin of this extra torque can be interpreted as the torque produced by the emission of angular momentum in the opposite direction, so that the radiated angular momentum is
\[
d{\bi S}_R=\gamma(v)[(d{\bi r}-d{\bi q})\cdot{\bi F}]{\bi S}_{CC},
\]
in the direction orthogonal to the zitter plane, i.e., the plane subtended by the two orthogonal vectors $(d{\bi r}-d{\bi q})$ and ${\bi u}$, or along the ${\bi S}_{CC}$ spin or in the opposite direction, depending on the sign of the difference of works of the Lorentz force $(d{\bi r}-d{\bi q})\cdot{\bi F}$. An effective radiation of this magnitude will produce as a reaction, the mentioned torque ${\bi M}_S$.

As a summary, the conservation of the mass $m$ and the absolute value of the spin for the center of mass observer ${\bi S}_{CM}(0)=1/2$, has lead to the existence of a reaction force ${\bi F}_R$ and a spin reaction torque ${\bi M}_S$, such that the particle radiates energy, linear momentum and angular momentum to the field in a time $dt$, given by:
\begin{eqnarray}
dH_R&=&|\gamma(v)[(d{\bi r}-d{\bi q})\cdot{\bi F}]H|,\label{difH}\\
d{\bi p}_R&=&\gamma(v)[(d{\bi r}-d{\bi q})\cdot{\bi F}]{\bi p},\label{difp}\\
d{\bi S}_R&=&\gamma(v)[(d{\bi r}-d{\bi q})\cdot{\bi F}]{\bi S}_{CC},\label{difSCM}
\end{eqnarray}
proportional, respectively, to the energy $H$, to the linear momentum ${\bi p}$ and to the CC spin ${\bi S}_{CC}$, multiplied by the same radiation factor $\gamma(v)[(d{\bi r}-d{\bi q})\cdot{\bi F}]$, which represents the difference of the work of the Lorentz force along the CC and CM trajectories, multiplied by the gamma factor $\gamma(v)$. When this work is positive the emission of positive energy is in the direction of the linear momentum ${\bi p}$, but if this work is negative this corresponds to the emission of positive energy, but in the opposite direction of ${\bi p}$. This is why we consider that the expression of the energy radiated corresponds to take the absolute value of this expression (\ref{difH}).

For the emission of linear and angular momentum, because they are vector magnitudes, they are emitted along the instantaneous ${\bi p}$ and ${\bi S}_{CC}$, in two possible directions according to the sign of the $(d{\bi r}-d{\bi q})\cdot{\bi F}$ ter. We have to compute the integration by the addition of the corresponding infinitesimal vector magnitudes.
This is a continuous radiation phenomenom between the particle and the field.

%%%%%%%%%%%%%%%%%%%%%
\section{Classical description of a photon}
\label{classphoton}
%%%%%%%%%%%%%%%%%%%%%
The model of a classical spinning photon is given in \cite{Bibphoton}. It is considered as a mechanical system of 6 degrees of freedom, where 3 represent the location ${\bi r}$ of the particle and another 3 the orientation of a comoving Cartesian frame attached to the point ${\bi r}$, that rotates with angular velocity $\bomega$. The free Lagrangian of the Dirac particle is a function $L_0({\bi u},{\bi a},\bomega)$ of the velocity and acceleration of the point ${\bi r}$ and on the angular velocity. 
The difference in this case is that the acceleration of the photon vanishes and the free Lagrangian of the photon is not a function of the acceleration. In this model we assume that the photon moves along a straight line with a constant velocity ${\bi u}=d{\bi r}/dt$, $u=c$ and rotates with an angular velocity $\bomega$.

The Lagrangian of a photon $L_0({\bi u},\bomega)$, in the International System of units is \cite{Bibphoton}
\[
L_0=\epsilon\frac{S}{c}{\bi u}\cdot\bomega,
\]
where $ \epsilon=\pm1$, represents the helicity, $c$ is the speed of light and the parameter $S$ is the absolute value of the spin. The canonical analysis leads to:
the linear momentum ${\bi p}=\partial L_0/\partial{\bi u}=\epsilon S \bomega/c$ and it is shown in that reference \cite{Bibphoton} that also has the same direction than the velocity ${\bi u}$.
The spin ${\bi S}=\partial L_0/\partial\bomega=\epsilon S{\bi u}/c$, and is not transversal. It has the direction of the velocity ${\bi u}$ for $\epsilon=1$ and the opposite direction if $\epsilon=-1$.  The parameter $S$ represents the absolute value of the spin and takes the same value in all inertial reference frames. All four vectors, ${\bi p}$, ${\bi u}$, ${\bi S}$ and $\bomega$ are collinear vectors. 
The Hamiltonian is defined as usual by
\[
 H={\bi p}\cdot{\bi u}+{\bi S}\cdot\bomega-L_0={\bi S}\cdot\bomega={\bi p}\cdot{\bi u}.
\]
The energy of the photon has two expressions $H={\bi p}\cdot{\bi u}=pc={\bi S}\cdot{\bomega}=S\omega$, so that the 
translation and rotation energy take the same value and represent the energy of the photon. A positive energy photon has a linear momentum along the velocity and the spin along the angular velocity, that can be righthanded or lefthanded. The four-momentum $p^\mu\equiv(H/c,{\bi p})$ and the invariant and constant of the motion $p^\mu p_\mu=0$ and the photon is a masless particle.
If $S=\hbar$ then $H=\hbar\omega=h\nu$, which is Planck's hypothesis concerning the quanta of the electromagnetic field. The photon is a boson. In natural units $H=\omega$, ${\bi p}=\epsilon\bomega$ and ${\bi S}=\epsilon{\bi u}$. The frequency of the classical photon corresponds to the frequency of its rotational motion around the direction of the motion. All vector magnitudes of the classical photon lie along the same straight line, the direction of its trajectory, as shown in the mentioned reference \cite{Bibphoton}.

%%%%%%%%%%%%%%%%%%%%%
\section{Conclusions about the emission of radiation}
\label{photon}
%%%%%%%%%%%%%%%%%%%%%

The radiation is related to the existence of two different characteristic points for the Dirac particle: the CC ${\bi r}$ and the CM ${\bi q}$. These two points are related to two different physical properties of matter: interaction and inertia, respectively. If they were the same point, like in the point particle case, radiation would not exist.

The continuous variation of the above observables (\ref{difH})-(\ref{difSCM}) depends on the values of the corresponding observables multiplied by difference of the work of the external force along the CC and CM, $(d{\bi r}-d{\bi q})\cdot{\bi F}$ and depends on the CM velocity through the gamma factor $\gamma(v)$. The continuous emission of linear momentum $d{\bi p}_R$ is along the direction of the linear momentum ${\bi p}$ at that time, forward or backward and the continuous emission of angular momentum $d{\bi S}_R$ takes place along the CC spin ${\bi S}_{CC}$ or in the opposite direction depending on the sign of the term $(d{\bi r}-d{\bi q})\cdot{\bi F}$. The energy of the radiated photon is the positive integral of that expression during a time $T$.

For the center of mass observer, ${\bi v}=0$, $d{\bi q}^*=0$ and the term $d{\bi r}^*\cdot{\bi F}^*=0$, because the CC velocity moves in the orthogonal direction to the external force and all radiations (\ref{difH})-(\ref{difSCM}) are zero as it corresponds to an observer who sees no changes in any of the intrinsic properties of the elementary particle. Otherwise, if the inertial observer at rest detects energy emission, this implies that the mass of the particle must decrease for this observer, which is contradictory to the atomic hypothesis. The inertial observer at rest with respect to the CM of the particle, does not detect radiation until the CM of the particle moves, and therefore from this later time onwards ceases to be the observer of the center of mass. To detect radiation in any reference frame, in addition to measuring the existence of an external force that produces acceleration of the CM, it is necessary that the CM moves in that reference frame. An electron at rest does not radiate and needs a finite acceleration time to reach a finite velocity, to emit a quantum of angular momentum 1.

The electromagnetism is a theory of continuous media. It was Planck who suggested that the transmission of radiation energy from an electromagnetic field was in the form of quanta of energy $h\nu$, where $\nu$ is the frequency of the radiated field and $h$ is Planck's constant. If we consider that the particle emits a photon, the intrinsic characteristics of a photon, as we have seen, are that it is a massless particle of spin $\hbar=1$, in natural units. In the above equations we have computed a continuous emission of energy, linear and angular momentum that are not grouped in the form of a quantum of electromagnetic energy of spin 1.

There are mathematical dificulties in describing radiation as the emission of discrete packets of energy, because the classical particle physics and field dynamics are mathematical formulations based on differential equations that involve the continuity of the functions and their derivatives. Planck's hypothesis that radiation occurs in the form of quanta of energy, introduces a discontinuity in the derivatives of the variables.
We have concluded that the non-modification of the intrinsic parameters of an elementary particle, mass and spin, as required by the atomic principle, implies the continuous transfer of energy, linear momentum and angular momentum between the particle and the field. Planck's hypothesis implies that angular momentum transfer is in the form of quanta of value $\hbar$. Radiation signifies the discontinuous emission of massless spinning particles of spin 1.

This analysis suggests that for spinless particles, where both the CC and CM are the same point, this difference in the work of the external field $(d{\bi r}-d{\bi q})\cdot{\bi F}=0$, cancels out and therefore charged spinless particles do not radiate. In Nature, there are no charged spinless elementary particles. Radiation is restricted to elementary spinning particles and is related to the existence of two different points ${\bi r}$ and ${\bi q}$, an external force and thus an acceleration of the CM, and to the resistance of mass and spin to being modified by the external fields.

We will say that a photon has been emitted if after a time $T$ the total angular momentum radiated continuously is equal to $\hbar$, or in natural units
\[
\Big|\int_t^{t+T} \;{d{\bi S}_{CMR}} \Big|=1.
\]
If the emitted particle is a photon, then the total energy and total linear momentum radiated during a time $T$
\[
H_R(T)=\int_t^{t+T} dH_R,\quad {\bi p}_R(T)=\int_t^{t+T} d{\bi p}_R,
\]
must necessarily satisfy the condition $H_R(T)^2={\bi p}_R(T)^2$, in natural units, because the photon is massless. 
The non-transverse nature of the photon spin implies that this radiated linear momentum  must be parallel or antiparallel to the radiated spin and therefore ${\bi p}_R(T)\sim \epsilon{\bi S}_{CMR}(T)$, where $\epsilon=\pm1$ is the helicity of the photon, and will probably depend on the particle spin orientation.
The quotient $\nu=H_R(T)/h$ is the frequency $\nu$ of the radiated photon, which from a classical point of view represents the frequency of its rotational motion.

In the dynamical equations (\ref{radreacCondi2}) all variables are continuous and differentiable.
If radiation exists, the dynamics must be discrete, meaning that the radiation does not occur continuously, but abruptly over time intervals of the order of $T$, when the angular momentum emission reaches the value 1. At that moment the linear momentum of the particle also changes abruptly by the magnitude ${\bi p}_R(T)$. 

The integration of the dynamic equations with radiation reaction will be a continuous integration without radiation until time $T$, when the emitted spin is 1. We stop the integration at that instant and change the boundary conditions of the CM velocity to the new one. We assume that at that time the linear momentum is ${\bi p}'(t)={\bi p}(t)-{\bi p}_R(T)$, and we beging the continuous integration again until the next emission of a new photon.
Since ${\bi p}'(t)=\gamma(v'){\bi v}'(t)$, $\gamma(v')=\sqrt{1+{p'}^2}$, the new CM velocity is ${\bi v}'(t)={\bi p}'(t)/\gamma(v')$, and we begin continuous integration again.

The trajectory of the CM will be a continuous trajectory but with a non-continuous derivative at the emission points. This trajectory has to be computed by using the equation (\ref{radreacCondi2}). The reason is that equation (\ref{radreacCondi2}) is obtained by spin conservation and it is the use of this modified equation that leads to the emission of energy, linear momentum and angular momentum. The CM velocity ${\bi v}$ changes abruptly at each instant of the emission and remains continuous and differentiable until the next point of emission. We will assume that, as far as the CC is concerned, the function ${\bi r}$ is also continuous and its time derivative ${\bi u}$ is discontinuous at the emission points.

The calculation of this radiation under an external electromagnetic field in the form of a spin-1 quantum  and the alignment betwen the radiated spin and the radiated linear momentum is left to a future article in which we will analyze both dynamic equations and the continuous zigzag trajectory of the particle's center of mass. We shall also analyze the stability of the nonlinear dynamic equations and the possible existence of some limit fields.

Preliminary numerical calculations of the radiated magnitudes imply that to emit a quantum of angular momentum $\hbar$ the integration has to be extended to very long integration times, of the order of $10^8$ to $10^{12}$ turns of the motion of the CC around the CM. Since in 1 second the CC of the electron gives $10^{21}$ turns, the above preliminary calculations imply that an electron in 1 second emits between $10^{9}$ to $10^{13}$ photons.

One prediction of the formalism is that a single electron under a uniform electric field, with the spin oriented in the direction of the field, does not radiate. All the energy expended by the field is transformed into mechanical energy of the electron. When we work with electrons, we interact with a large number of them in the form of a beam or a current. As shown in \cite{2elec} a very close electron-electron interaction modifies the spin orientation of both particles, and therefore the electrons in a beam are continuously varying the orientations of their spins due to mutual interactions. It seems dificult to control that a  beam of electrons accelerated in a uniform electric field does not radiate, but if we are able to control the orientations of their spins, the intensity of the radiation would be dampened.

%%%%%%%%%%%%%%%%%%%%
\ack{I want to thank Oliver Consa, founder of the Zitter Institute, for pointing out the relationship between the $K_E$ and $K_B$ conversion factors of equation (\ref{natdifeq}) and the so-called Schwinger limits for the electric and magnetic fields, respectively.}

%%%%%%%%%%%%%%%%%%%%%%%%%%%%%

\end{document}